\documentclass[aps,twocolumn,prb]{revtex4-1}
\usepackage{comment}
\usepackage{graphicx}
\usepackage{dcolumn}
\usepackage{bm}
\usepackage{tabularx}
\usepackage{array}

\newcolumntype{P}[1]{>{\centering\arraybackslash}p{#1}}

\begin{document}

\author{A.~C. McRae, V. Tayari, J.~M. Porter, and A.~R. Champagne}
\email{a.champagne@concordia.ca}
\affiliation{Department of Physics, Concordia University, Montr\'{e}al, Qu\'{e}bec, H4B 1R6, Canada}
\title{Giant electron-hole charging energy asymmetry in ultra-short carbon nanotubes}

\begin{abstract}
Making full usage of bipolar transport in single-wall carbon nanotube (SWCNT) transistors could permit the development of two-in-one quantum devices with ultra-short channels. We report on clean $\sim$10 to 100 nm long suspended SWCNT transistors which display a large electron-hole transport asymmetry. The devices consist of naked SWCNT channels contacted with sections of SWCNT-under-annealed-gold. The annealed gold acts as an n-doping top gate which creates nm-sharp barriers at the junctions between the contacts and naked channel. These tunnel barriers define a single quantum dot (QD) whose charging energies to add an electron or a hole are vastly different ($e-h$ charging energy asymmetry). We parameterize the $e-h$ transport asymmetry by the ratio of the hole and electron charging energies $\eta_{e-h}$. We show that this asymmetry is maximized for short channels and small band gap SWCNTs. In a small band gap SWCNT device, we demonstrate the fabrication of a two-in-one quantum device acting as a QD for holes, and a much longer quantum bus for electrons. In a 14 nm long channel, $\eta_{e-h}$ reaches up to 2.6 for a device with a band gap of 270 meV. This strong $e-h$ transport asymmetry survives even at room temperature.
\end{abstract}

\maketitle
A unique feature of ultra-low disorder SWCNT channels is that a small gate voltage, $V_{G}$, can tune them from n-type (electron-doped) to p-type (hole-doped) devices. In principle, this feature could allow the fabrication of two-in-one transistors with one set of characteristics for electron transport and another for hole transport. Unfortunately, the intrinsic transport properties of SWCNTs are mostly electron-hole ($e-h$) symmetric\cite{JarilloHerrero04,Laird15}, and thus the two types of transport are redundant. Nevertheless, SWCNT devices can be engineered to create an $e-h$ transport asymmetry. This is achieved for instance in SWCNT quantum dot (QD) transistors whose tunnel barrier heights depend on whether the channel is n or p doped. Such a transport asymmetry has been demonstrated in ultra-clean devices with channel lengths ranging from a few hundreds to $\sim$ 100 nanometers \cite{Steele09, Jung13}. Downsizing $e-h$ asymmetric SWCNT transistors to $\sim10$ nm would benefit both applied research to develop commercial SWCNT transistors\cite{Franklin12,Shulaker13}, and fundamental research to make use of spin\cite{Kuemmeth08,Pei12, Laird13, Laird15} and mechanics\cite{Island12, Chaste12, Benyamini14, Moser14, Weber15} in quantum nano-electro-mechanical systems (NEMS). For example, such SWCNT transistors could scale down the size of SWCNT qubits\cite{Wunsch09, Pei12}, enable the study of quantum systems with both strong electron-phonon and electron-electron interactions\cite{Cornaglia04}, and make devices which can switch from being QDs (0D) to quantum buses (1D).

Ultra-clean SWCNTs used to explore many-body physics and quantum bits have previously been limited to channel lengths above $\sim$ 100 nm \cite{Wu10,Pei12, Laird13, Benyamini14,Cao04,Leroy04,Kuemmeth08, Steele09}. Here, we demonstrate the capability to engineer ultra-clean (suspended) SWCNT-QD transistors with a giant $e-h$ charging energy asymmetry and channel lengths down to 14 nm. The key feature of our fabrication is to use an annealed-gold film as an electrostatic gate directly deposited on a SWCNT (no dielectric spacer). This n-dopes the gold-covered SWCNT sections which then act as contacts to the naked channel. Theses gold gates create extremely sharp barriers at the contact-to-channel interfaces ($\approx$ 3 nm wide), and these barriers have different heights whether the channel is n or p doped. We measured transport in five devices under both electron and hole doping of their channels, and at temperatures ranging from 1.3 to 295 K. In Table 1, we list each device's bang gap $E_{g}$ as extracted via QD transport, channel length $L_{SEM}$ measured via SEM imaging, channel length $L_{G}$ extracted from QD transport data, and ratio $\eta_{e-h}$ of the charging energies for holes, $E_{C}^{h}$, and electrons, $E_{C}^{e}$. In a small gap SWCNT device (Device A), $E_{C}$'s for electrons and holes differ by orders of magnitude. When this device's channel is p-doped, it forms a 102 nm-long low-disorder QD. When the channel is n-doped, the device becomes a one-dimensional waveguide where carriers can travel $\approx$ 330 nm without losing their phase coherence. In the four devices whose band gaps $E_{g}$ $\gtrsim$ 200 meV the charging energy asymmetry ratio $\eta_{e-h}$ ranges from 1.5 to 2.6. We show that the $e-h$ charging asymmetry scales inversely with channel length and the SWCNT's band gap. The $E_{C}^{h}$ reaches up to 100 meV in our shortest device (Device B), demonstrating the potential for room temperature operation.
\begin{table*}
\centering
\begin{tabular}{|P{2.5cm}|P{2.5cm}|P{2.5cm}|P{2.5cm}|P{2.5cm}|}
\hline
Device &$E_{g}$ (meV) & $L_{SEM}$ (nm) & $L_{G}$ (nm) & $\eta_{e-h}$\\
\hline
A&28 $\pm$ 5 & 111 $\pm$ 5 & 102 $\pm$ 5 &$\gtrsim$ 100 \\
B&270 $\pm$ 50 & 14 $\pm$ 3 & 7 $\pm$ 5 &2.6\\
C&190 $\pm$ 20 & 42 $\pm$ 7 & 46 $\pm$ 8 &1.5\\
D&250 $\pm$ 20 & 16 $\pm$ 4 & 13 $\pm$ 5 &2.1\\
E&170 $\pm$ 50 & 24 $\pm$ 8 & 15 $\pm$ 5 &2.5\\
\hline
\end{tabular}
\newline\newline
\caption{\label{} Key parameters for the five SWCNT-QD transistors reported. $E_{g}$ is the band gap extracted from transport data, $L_{SEM}$ the channel length measured via SEM, $L_{G}$ the channel length from transport data, and $\eta_{e-h}= E_{C}^{h} / E_{C}^{e}$ is the measured charging energy asymmetry between electron and hole doping of the quantum dots.}
\end{table*}

\begin{figure*}
\includegraphics[width=6.25in]{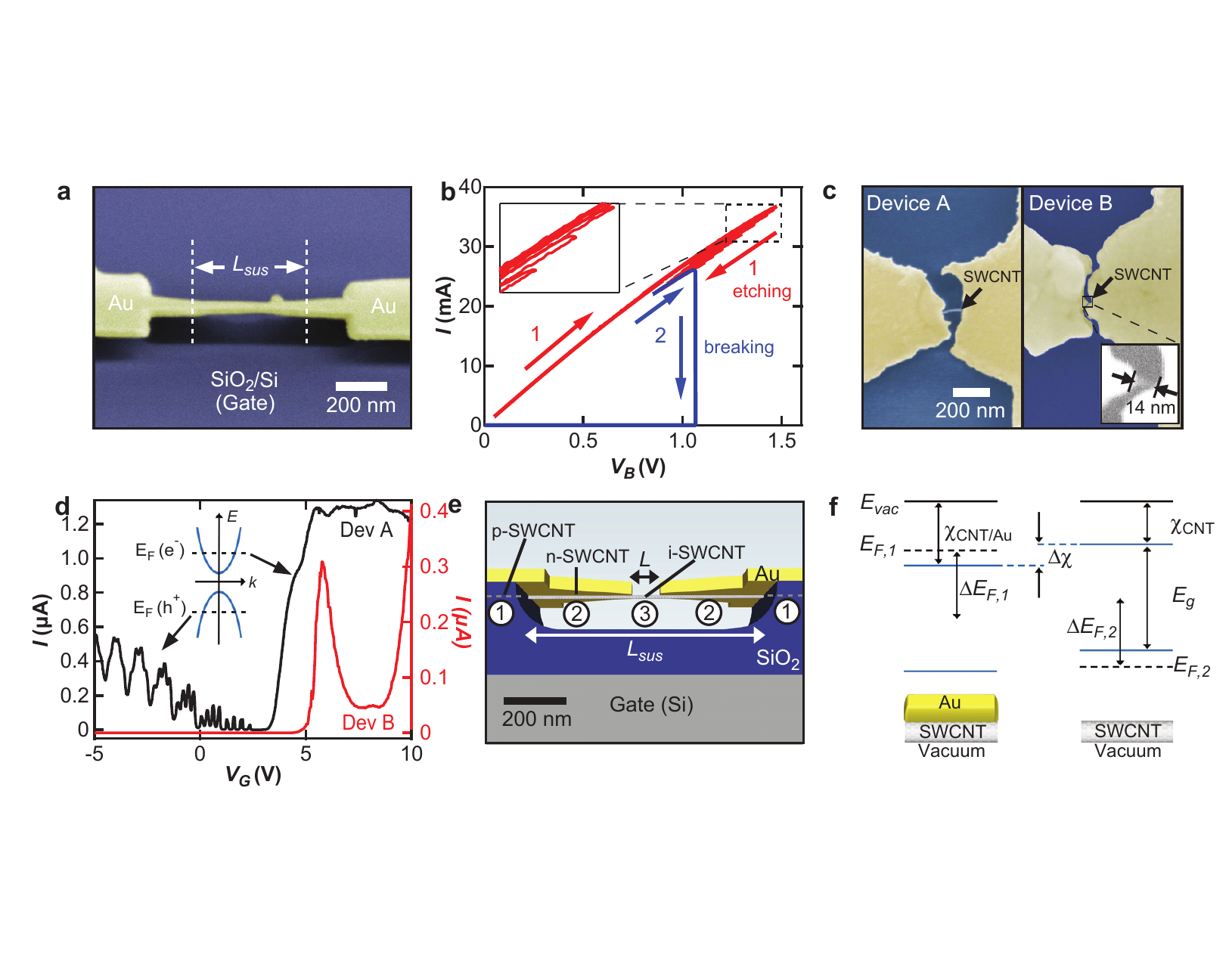}
\caption{\label{}Ultra-short suspended SWCNT-QD transistors. \textbf{a}, Tilted scanning electron microscope (SEM) image showing a suspended gold breakjunction fabricated on top of a SWCNT. The breakjunction is suspended over a length $L_{sus} \approx$ 350 nm. The back plane (blue) is used as a global back-gate. \textbf{b}, The $I-V_{B}$ electromigration (EM) data for a gold-on-SWCNT breakjunction at $T =$ 4 K, showing the process to (1) narrow the junction and (2) create a naked SWCNT channel. \textbf{c}, SEM images of Devices A and B after EM. The naked channels are visible and their lengths are $L_{SEM} =$  111 and 14 nm, respectively. \textbf{d}, $I-V_{G}$ transistor data from Devices A (black) and B (red) at $T=$ 1.3 K. We measured a much higher conductance for positive $V_{G}$. This indicates that the suspended gold, annealed during the EM, n-dopes the underlying SWCNT. \textbf{e}, Geometry of our suspended SWCNT transistors. The labels 1, 2, and 3 refer respectively to the p-doped on-substrate nanotube sections, the n-doped gold-covered suspended tube sections, and the naked SWCNT channel. \textbf{f}, The electronic bands in the gold-covered (left) and naked (right) suspended tube sections. The quantity $E_{g}$ is the band gap. The quantities $E_{F,1}$, $E_{F,2}$, $\Delta E_{F,1}$, $\Delta E_{F,2}$ are the Fermi energies and the Fermi energy shifts from the center of the band gaps in the gold-covered and naked SWCNT, respectively. $\chi_{CNT/Au}$ and $\chi_{CNT}$ are the electron affinities in the gold-covered and naked sections, while $\Delta \chi$ is the electron affinity difference between the two.}
\end{figure*}

\subsection{Ultra-short suspended SWCNT-QD transistors with ballistic contacts}
Figure 1 summarizes the fabrication and contact geometry of our suspended SWCNT devices\cite{Island11, Island12}. We first fabricated suspended gold-on-SWCNT breakjunctions (see Methods). Figure 1a shows a breakjunction and $L_{sus}$ indicates the length over which the gold is suspended. The substrate (blue) is used as a global back-gate electrode. The final fabrication step is to create a nm-long naked SWCNT channel in the center of the gold breakjunction. To do so, we used a previously reported feedback-controlled electromigration (EM) procedure\cite{Island12,Tayari15} summarized in Fig.\ 1b (see Methods). This EM process exposes a short naked SWCNT channel. Figure 1c shows top-view SEM images of Devices A and B whose channel lengths are respectively 111 $\pm$ 5 nm and 14 $\pm$ 3 nm. Figure 1d shows the $I-V_{G}$ characteristics at $T =$ 1.3 K for Devices A (black) and B (red). The EM not only uncovers short naked SWCNT channels, but also anneals them such that the QDs in the channels are nearly undoped\cite{Island12} with their charge neutrality point (CNP) close to $V_{G} = $ 0. Close inspection of the gold surface (Fig.\ S1) reveals that the suspended portion of the gold film changes texture after the EM. The inset of Fig.\ 1c shows the Fermi level position in the channel under electron or hole doping. The current (conductance) is much higher when the back gate n-dopes rather than p-dopes the channel. This is consistent with the suspended gold film (top gate) n-doping the SWCNT sections it covers, and that these SWCNT-under-gold sections act as contacts to the channel (see Fig.\ 1e). We emphasize that we use pure gold films, without any adhesion layer, to create clean SWCNT-Au interfaces. Gold's workfunction is very close to the threshold, $\approx$ 5.4 eV, where physisorbed metal films switch from n-doping to p-doping graphene \cite{Giovannetti08} and SWCNTs\cite{Cui03,Hasegawa11}. A small change in gold's workfunction due to the removing of oxygen adsorbed on the gold during the EM can tune the gold films from p to n doping \cite{Heinze02,Cui03}. We observed this effect in Device B (Fig.\ S2). We note that the fabrication process of our ultra-clean SWCNT devices is compatible with the optical measurement of the SWCNT chirality \cite{Jorio01, Liu13} prior to the gold deposition. Such ultra-clean SWCNT transistors with a known tube chirality could lead to breakthroughs in SWCNT physics \cite{Laird15}.

The electronic interactions between an ultra-clean SWCNT and gold are weak\cite{Knoch08} and the density of states in the SWCNT is unaffected by the gold film. This is because SWCNT electronic wavefunctions have large wavevectors, and the conservation of momentum suppresses the tunneling matrix elements between these high-momentum states and gold states\cite{Tersoff99}. It follows that the injection length for an electron to tunnel between gold and the SWCNT is very long ($\sim \mu m$) \cite{Knoch08, Sundaram11}. Electrostatic disorder leads to stronger SWCNT-gold interactions and shorter injection lengths \cite{Tersoff99,Knoch08}. This knowledge is necessary to describe the electron transport in the gold covered sections of our SWCNTs, and the device design shown in Fig.\ 1e. The gold film covering the SWCNT sections \textcircled{\raisebox{-.9pt} {1}} is thermally anchored to the substrate, and not annealed during the EM process. This leaves the sections \textcircled{\raisebox{-.9pt} {1}} p-doped and acting as diffusive (disordered) contacts. The suspended gold film covering the tube sections labeled \textcircled{\raisebox{-.9pt} {2}} is annealed as evidenced from both transport data (Fig.\ 1d) and the film texture (Fig.\ S1). Finally, the naked SWCNT channel labelled \textcircled{\raisebox{-.9pt} {3}} is thoroughly annealed by the EM process \cite{Island11, Island12} and can be doped with either holes or electrons via $V_{G}$. The data presented below will demonstrate that the charge transport is ballistic in both sections \textcircled{\raisebox{-.9pt} {2}} and \textcircled{\raisebox{-.9pt} {3}} of the SWCNTs, and that $e-h$ asymmetric tunnel barriers form where the two sections connect.

The left hand side of Fig.\ 1f shows an example of the band diagram for the SWCNT contacts (sections \textcircled{\raisebox{-.9pt} {2}}), and the right hand side shows the bands in the naked channel (section \textcircled{\raisebox{-.9pt} {3}}). The dashed lines indicate the positions of the Fermi energies $E_{F,1}$ and $E_{F,2}$ in each section, and their shifts $\Delta E_{F,1}$ and $\Delta E_{F,2}$ away from the center of the band gap. The band gap has the same value in both SWCNT sections since they belong to the same tube. The nanotube's electron affinity, $\chi_{CNT}$, is defined as the energy between the bottom of the conduction band and the vacuum energy, $E_{vac}$. On the left side of Fig.\ 1f, gold transfers charges to the SWCNT and moves $E_{F,1}$ up from the center of the gap. The exact amount of doping varies depending on the crystalline orientation of the gold as well as the quality of annealing (oxygen content). The relevant values of $\Delta E_{F,1}$ for our devices are reported to range from 0.05 to 0.2 eV \cite{Cui03, Khomyakov09, Sundaram11, Hasegawa11, Chaves14}, and we use a median value of 0.12 eV to draw our band diagrams. The gold film can also modify the nanotube's electron affinity by $\Delta \chi$. This shift is expected to be 0.03 - 0.05 eV \cite{Khomyakov09,Hasegawa11}, and we use 0.05 eV to draw the bands in our devices. These parameters correctly predict several features of the data below, such as the presence or absence of a barrier, the sign of the $e-h$ charging energy asymmetry, and correlate with the magnitude of the asymmetry. Moreover, the discussion and conclusions below remain valid over the range of reported $\Delta E_{F,1}$ and $\Delta \chi$ values. The Fermi energy of the naked SWCNT channel, $E_{F,2}$, can be tuned using $V_{G}$. The gating efficiency of the back gate (dielectric of 130 nm vacuum plus 170 nm SiO$_{2}$) is much weaker than the gating efficiency of the gold film (top gate) which is only three angstroms away from the SWCNT \cite{Hasegawa11}. This means that the back gate does not significantly affect $E_{F,1}$, but only effectively tunes $E_{F,2}$. This is confirmed by the transport data below showing that the lengths (confinement) of the SWCNT-QDs are independent of $V_{G}$. At the junctions between the tube sections \textcircled{\raisebox{-.9pt} {2}} and \textcircled{\raisebox{-.9pt} {3}}, the band diagrams in Fig.\ 1f are brought into contact and equilibrate to form homojunctions.

\begin{figure*}
\includegraphics[width=6.0in]{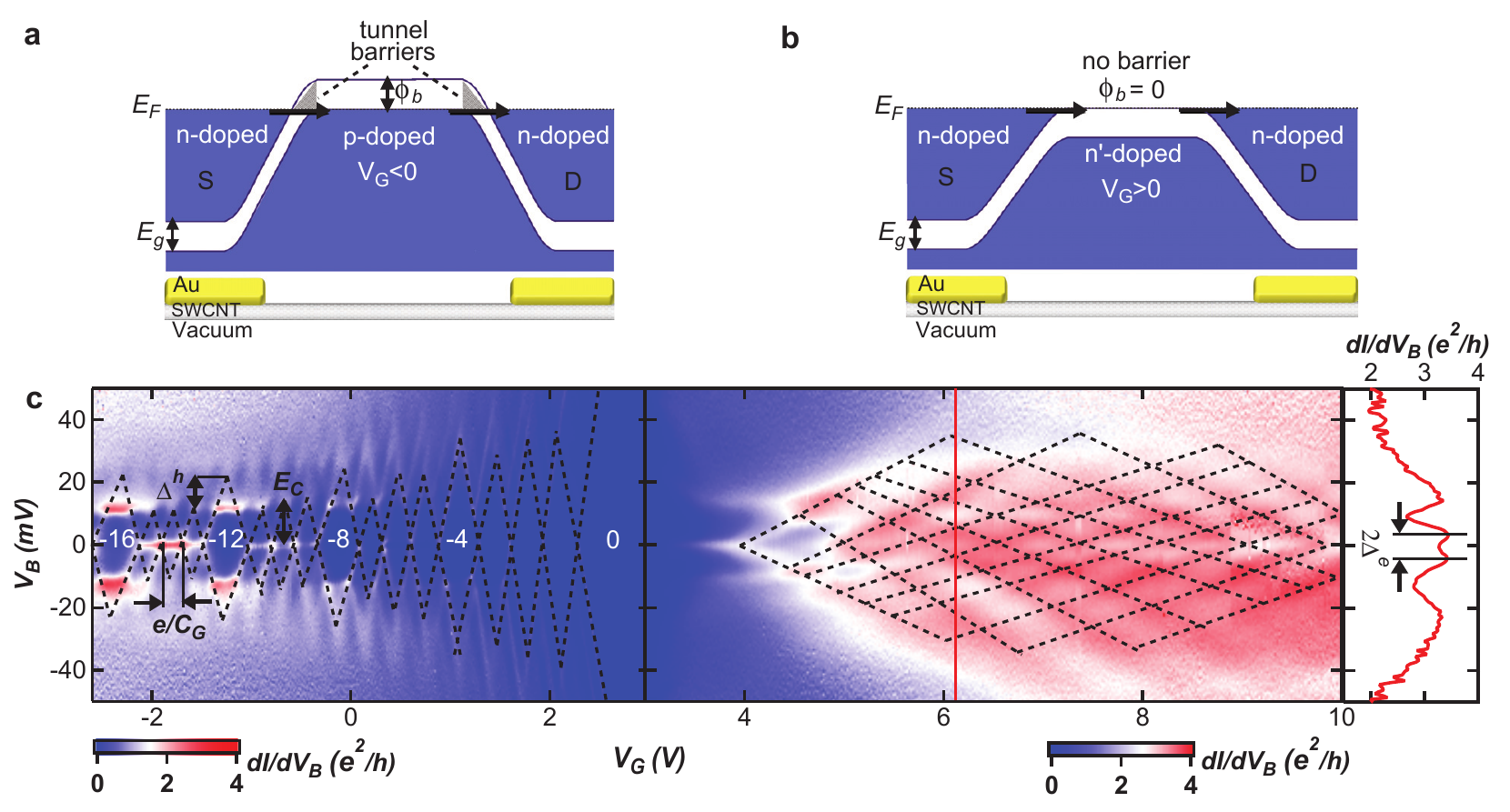}
\caption{\label{}Electron-hole transport asymmetry and ballistic contacts. \textbf{a-b}, Schematics of the bands at the junctions between the n-doped (gold-covered) SWCNT contacts and naked channel in Device A ($E_{g} =$ 28 meV) when the channel is \textbf{a} p-doped and \textbf{b} n-doped. Band-to-band tunnel barriers (shaded triangles) of height $\phi_{B} = E_{g}$ form between the SWCNT contacts and the p-doped channel. On the other hand, when the naked channel is n-doped $\phi_{B} =$ 0. \textbf{c}, Charge transport data ($dI/dV_{B}-V_{B}-V_{G}$) for Device A at $T=$ 1.3 K. The charge neutrality point of the channel is clearly visible around $V_{G}$ = 3 V. A striking asymmetry is visible in the transport data between hole doping ($V_{G} < 3$ V) and electron doping ($V_{G} > 3$ V) of the channel. For hole doping, a four-fold Coulomb diamond structure indicates a single clean quantum dot (QD). The gate-to-QD capacitance $C_{G}$ corresponds to a channel length of 102 $\pm$ 5 nm, closely matching $L_{SEM}$ in Fig.\ 1c. Under electron-doping, the transport data show clear Fabry-P\'{e}rot quantum interferences (see vertical outset). The spacing between conductance maxima in $V_{B}$ is 5 $\pm$ 1 mV, giving $L_{FP} =$ 330 $\pm$ 70 nm. This matches the suspension length of the gold film $L_{sus}=$ 350 $\pm$ 70 nm (Fig.\ S1). It confirms that the transport in the gold-covered nanotube sections is ballistic and preserves the quantum phase of the electrons traveling through the channel.}
\end{figure*}

Figure 2a,b show schematics of the homojunctions (SI section 4) in Device A, $E_{g} =$ 28 meV. Figure 2a shows the band alignment when the channel is hole doped, while Fig.\ 2b shows the electron doping configuration. Tunnel barriers form at the contacts when the channel is p-doped. The transport mechanism across the barriers in Fig.\ 2a is band-to-band tunneling, and has been previously described for SWCNT transistors \cite{Appenzeller04, Steele09}. Approximating the shape of the bands at the homojunctions using the WKB model\cite{Schmidt13} (shaded triangular regions), the tunnel barrier height is $\phi_{B} = E_{g} = $ 28 meV. When the channel is electron doped (Fig.\ 2b), the charge transport only involves the conduction band, and there is no tunnel barrier. Low-temperature ($T =$ 1.3 K) transport data from Device A are shown in Fig.\ 2c, where the color scales shows the differential conductance $dI/dV_{B}$ as a function of $V_{G}$ and bias voltage $V_{B}$. The negative (positive) integer labels $N$ indicate the number of holes (electrons) in the SWCNT channel. Under hole-doping of the channel, clear Coulomb blockade diamonds indicate quantum dot transport. When the channel is electron-doped, transport data along the $V_{G}$ and $V_{B}$ directions show strong Fabry-P\'{e}rot interferences characteristic of ballistic transport \cite{Liang01, Jorgensen09}. The dramatic asymmetry in quantum transport between holes (quantum dot) and electrons (ballistic channel) is a consequence of the different contact barriers for the two types of doping. A closer look at the data in Fig.\ 2c will confirm the device geometry shown in Fig.\ 1e.

On the left hand side of Fig.\ 2c, we observe four-fold quantum degeneracy\cite{Laird15} of the hole-doped QD energy levels. We extract an energy level spacing for holes of $\Delta^{h} \approx$  12 $\pm$ 1 meV (height difference between neighbouring tall and short diamonds), and a charging energy $E_{C}^{h}$ = 11 $\pm$ 1 meV (height of short diamonds). The width of the Coulomb diamonds for holes is 0.21 $\pm$ 0.01 V $ = e/C_{G}$, where $C_{G}$ is the capacitance between the QD and back-gate. We obtain a QD length, $L_{G} =$ 102 $\pm$ 5 nm, from the measured $C_{G}$ using a wire over a plane capacitor model (SI section 3). This length closely matches the length of the channel as measured by SEM, $L_{SEM} =$ 111 $\pm$ 5 nm. We note that $L_{G} < L_{SEM}$ is expected due to the two finite width p-n or n-n' junctions at each end of the QD. Using the measurements of $L_{G}$ and $L_{SEM}$ in all devices (Table 1), we find an average p-n junction length of $L_{pn} =$ 3 $\pm$ 1 nm. An independent estimate of $L_{pn}$ is obtained from the SWCNT screening length\cite{Knoch08}, $\lambda = \sqrt{(\epsilon_{CNT}/\epsilon_{vac})d_{CNT}d_{vac}}$, where $\epsilon_{CNT} \approx$ 10 is the SWCNT permittivity\cite{Lu07}, $d_{CNT} \sim $ 1.3 nm is the tube diameter, $\epsilon_{vac} =$ 1 is the permittivity of vacuum, and $d_{vac} \approx$ 0.3 nm \cite{Nemec06,Hasegawa11} is the thickness of the vacuum dielectric between the tube and gold. This gives $\lambda \approx$ 2 nm, matching the $L_{pn}$ extracted above. It is remarkable that we can think of the suspended (annealed) gold film as a gate electrode for the SWCNT, with a dielectric spacer of $\sim$ 3 {\AA}, offering the possibility to create extremely sharp p-n junctions. This capability will find applications in testing the ultimate downscaling of SWCNT transistors \cite{Island11, Franklin12}, and creating nm-sized coherent electronic devices\cite{Kuemmeth10, Laird13,Pecker13,Laird15} and NEMS\cite{Island12,Chaste12,Benyamini14,Moser14, Weber15}.

While the suspended gold gates in Device A confine holes to a QD of length $L_{G} =$ 102 nm. The gold gates do not confine electrons inside the naked channel of Device A. Using the transport data in the vertical line cut of Fig.\ 2c, we measure an energy level spacing $\Delta = e\Delta V_{B}=$ 5 $\pm$ 1 meV between the interference maxima. We extract the length of the electron cavity as $L_{FP} = hv_{F}/(2\Delta) =$ 330 $\pm$ 70 nm, where $v_{F} = $ 8 $\times$ $10^5$ m/s is the Fermi velocity. This cavity length matches the suspension length of the gold gates, $L_{sus} \approx$ 350 $\pm$ 70 nm (Figs.\ 1e and S1). The coherent FP interferences and cavity length confirm that the majority of the current in the contact sections \textcircled{\raisebox{-.9pt} {2}} flows through the SWCNT and not the gold film. This is expected due to the long charge injection length in clean SWCNTS \cite{Nemec06, Knoch08, Sundaram11}. The FP interferences also imply that the annealed gold covering sections \textcircled{\raisebox{-.9pt} {2}} does not prevent ballistic transport in the underlying SWCNT\cite{Yang13}. Using an open QD model\cite{Jorgensen09}, we extract an approximate electron charging energy of $E_{C}^{e}\sim$ 0.1 meV, giving a charging energy asymmetry ratio $\eta_{e-h} = E_{C}^{h} / E_{C}^{e} \gtrsim$ 100. We note that previously reported SWCNT devices which showed QD to FP $e-h$ asymmetry \cite{Cao04,GroveRasmussen07,Makarovski07} were not fabricated with suspended and annealed gold films, and thus had equal channel lengths for holes and electrons. The important new feature in Fig.\ 2c is that annealed-suspended gold film acts as a local top gate which creates a two-in-one quantum device with different channel lengths for electron and hole transport. Specifically, applying a small $V_{G}$ to Device A toggles between a 102 nm QD and a 330 nm coherent wire. In the following section, we demonstrate that large $e-h$ transport asymmetries, and the ensuing dual device functionality, can also be achieved in shorter SWCNT-QD transistors whose $E_{g}$ ranges up to 270 meV.

\begin{figure*}
\includegraphics[width=6.25in]{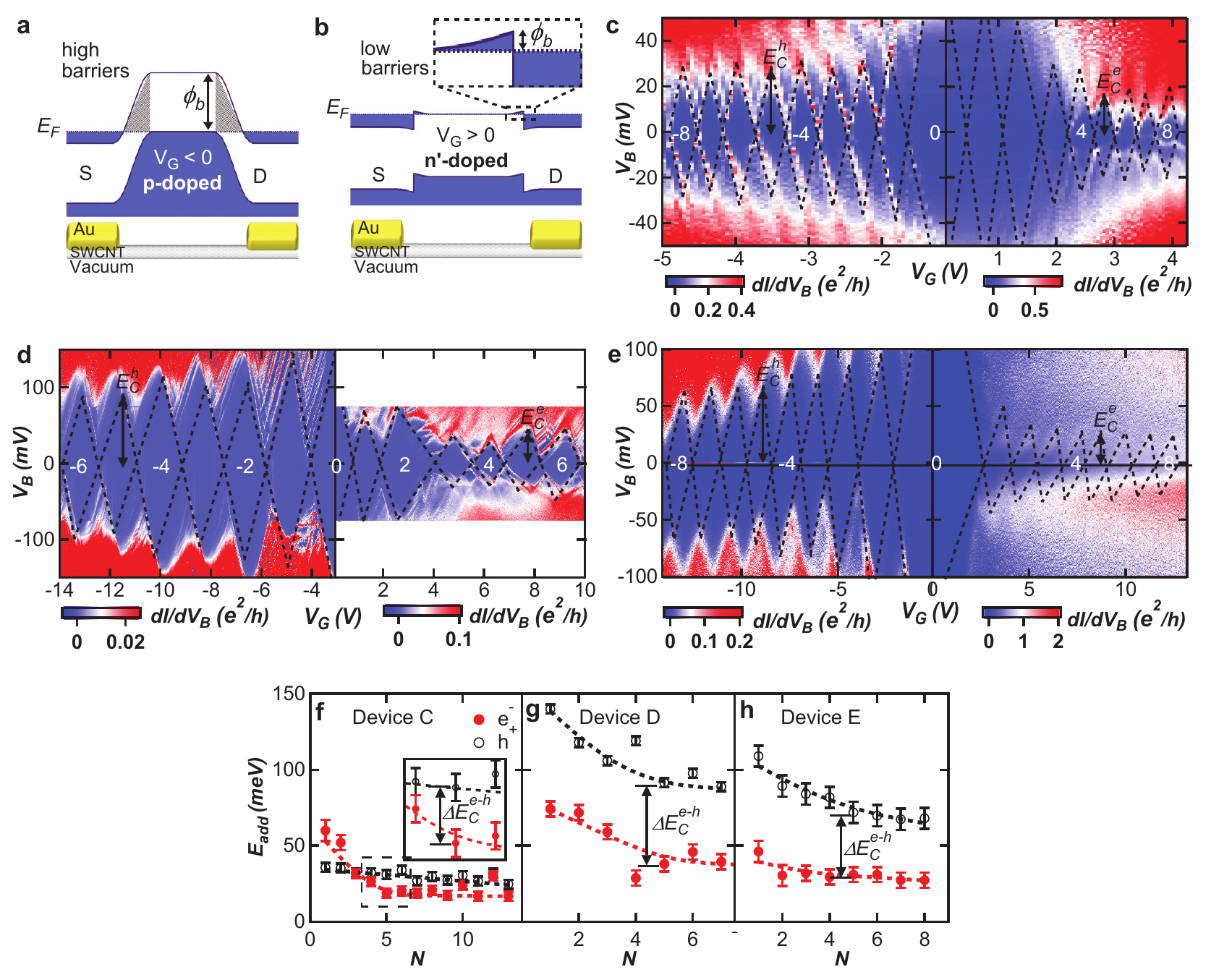}
\caption{\label{}Giant $e-h$ charging energy asymmetry in $\gtrsim$ 200 meV band gap SWCNT-QDs. \textbf{a-b}, Schematics of the junctions between the n-doped SWCNT contacts and naked SWCNT channel in Device C, $E_{g} =$ 190 meV. \textbf{a} When the channel is p-doped, the band-to-band tunnel barriers (shaded triangles) have a height $\phi_{B}=$ 190 meV. \textbf{b} When the channel is n-doped, the tunnel barriers are much smaller, $\phi_{B}\sim$ 12 meV (SI section 4). \textbf{c-e}, $dI/dV_{B}-V_{B}-V_{G}$ transport data for Devices C, D, and E respectively ($T$= 4.0 K, 1.3 K and 50 K). In all devices, clear Coulomb diamonds are visible for both hole and electron channel dopings, and show the formation of a single QD in the channel. The charging energies, i.e. heights of the odd-labeled diamonds, are much larger for holes than for electrons due to the tunnel barrier asymmetry. \textbf{f}, Extracted addition energy, $E_{add}$ = height of diamonds, versus charge number, $N$, for both holes (open black data) and electrons (filled red data) in Devices C, D, and E. The data for odd $N$, for which $E_{add} = E_{C}$, are interpolated with dashed lines. The charging energies decrease with increasing $N$, and a clear offset, $\Delta E_{C}^{e-h}$ is visible between hole and electron data. The value of $\Delta E_{C}^{e-h}$ becomes roughly constant at large $N$.}
\end{figure*}

\subsection{Giant $e-h$ charging energy asymmetry in $\gtrsim$ 200 meV band gap SWCNT-QDs}
For practical applications, we would like to extend the ability to create large $e-h$ transport asymmetry to SWCNT transistors whose $E_{g}$ are sufficiently large for room-temperature operation, or to act as two-in-one QD transistors. In all four large $E_{g}$ devices we studied, the channels host single QDs whether they are doped with holes or electrons. However, the heights of the contact barriers for the two doping configurations are vastly different.

Figure 3a,b show the homojunctions forming at the interfaces between the contacts and channel in Device C. When the channel is hole doped (Fig.\ 3a), tall tunnel barriers of $\phi_{B}\approx E_{g}=$ 190 meV form where the contacts (sections \textcircled{\raisebox{-.9pt} {2}}) meet the SWCNT channel (sections \textcircled{\raisebox{-.9pt} {3}}). Figure 3b shows the configuration when the channel is electron doped. The transport across the n-n' junctions only involves the conduction band, and the barriers are much smaller, $\phi_{B}\approx$ 12 meV. This $e-h$ barrier height asymmetry creates an $e-h$ asymmetry of the source-QD and drain-QD capacitances, $C_{S}$ and $C_{D}$ respectively. It follows that the QD charging energy $E_{C} = e/(C_{S}+C_{D}+C_{G}) = e/C_{\Sigma}$, depends on whether the QD is populated with holes or electrons. This is visible in Figs.\ 3c-e where $dI/dV_{B} - V_{B} - V_{G}$ data for Devices C, D and E are shown. The Coulomb diamond heights in Fig.\ 3c-e, i.e. the addition energies $E_{add} = E_{C} + \Delta$, are much larger for holes (left) than for electrons (right).

We use the data in Fig.\ 3 to quantify both the $e-h$ charging energy asymmetry $\Delta E_{C}^{e-h} = E_{C}^{h} - E_{C}^{e}$ and the asymmetry ratio $\eta_{e-h} = E_{C}^{h}/E_{C}^{e}$. We extract from Fig.\ 3c-e the heights of the diamonds as a function of the charge occupation number $N$ as shown in Fig.\ 3f-h. The dashed lines are interpolations of the odd-$N$ data points, for which $E_{add} = E_{C}$ since Devices C, D, and E show a two-fold degenerate energy spectrum. We note that $\Delta E_{C}^{e-h}$ converges to a roughly constant value at large $N$. To compare the asymmetry between devices, we extract $\eta_{e-h}$ at $N = 5$. The $\eta_{e-h}$ in Devices C, D and E are respectively 1.5, 2.1 and 2.5 (SI section 5). While the $E_{g}$ for all three of these devices are comparable, the channels in Devices D and E are three times shorter than in Device C. Both Devices D and E have significantly larger $\eta_{e-h}$ than Device C, and this correlation is also confirmed by Device B (SI section 2). We thus conclude that $\eta_{e-h}$ scales inversely with length. This explains why we can observe large $\eta_{e-h}$ in ultra-short QDs whose band gaps range up to 270 meV, while previous experiments\cite{JarilloHerrero04} on much longer QD devices (few 100s of nm) showed $\eta_{e-h}\approx$ 1 in devices with similar band gaps. To further understand the length dependence of $\eta_{e-h}$, it is useful to analyze the capacitances of the SWCNT-QDs versus $L_{G}$.

In Fig. 4a-b, the data in black (red) are for hole (electron)-doped channels. Figure 4a shows $C_G$ extracted from the widths of the $N =$ 5 Coulomb diamonds for all devices (except for Device B, where only $N=$ 1 is available) versus $L_{G}$. $C_{G}$ is the same for hole or electron occupations of the QD and ranges from $\approx$ 0.05 to 0.6 aF. As expected, $C_{G}$ scales linearly with $L_{G}$. Figure 4b shows the total QD capacitance $C_{\Sigma} = C_{S}+C_{D}+C_{G}$ extracted from the slopes of the diamonds in Figs.\ 2 and 3 as a function of $L_{G}$ (see SI section 2). There is a marked $e-h$ asymmetry between the total capacitances for holes and electrons, which we label $\Delta C^{e-h}_{\Sigma} = C_{\Sigma}^{e} - C_{\Sigma}^{h}$. Figure 4c shows that both the relative capacitance asymmetry $\Delta C^{e-h}_{\Sigma} / C_{\Sigma}^{h}$ (left axis) and the resulting $\eta_{e-h}$ (right axis) decrease rapidly with $L_{G}$ for Devices B, C, D, and E which all have similar band gaps.

We also observed the qualitative effect of $E_{g}$ on $\eta_{e-h}$. While a length change of a factor of three between Devices C, D, and E (similar $E_{g}$'s) changes $\eta_{e-h}$ by less than a factor of 2. An even smaller change of length between Devices A (Fig.\ 2c) and C (Fig.\ 3c), but coupled with an order of magnitude change in $E_{g}$, leads to a difference of two orders of magnitude in $\eta_{e-h}$. The change in $\eta_{e-h}$ between Devices A and C can thus be predominantly ascribed to their different $E_{g}$'s. This inverse dependence of $\eta_{e-h}$ on $E_{g}$ can be understood by inspecting Fig.\ 2a,b. The relative $e-h$ barrier height asymmetry decreases rapidly as $E_{g}$ increases and becomes larger than the doping $\Delta E_{F,1}$ induced by the gold gates.

Because the values of $E_{C}$ and $\phi_{B}$ in our SWCNT-QD transitors are large compared to $k_{B}T_{room}$, we can envision making use of the $e-h$ transport asymmetry in room temperature devices. Figure 4d shows $I - V_{G}$ data for Device B at 295 K. The SEM image of Device B in Fig.\ 1c shows a channel length of only 14 nm, and data taken in the same device show a bandgap of 270 meV (SI section 2). During a warm up from 1.3 K to room temperature, Device B was exposed to oxygen (SI section 1). This exposure changed the suspended gold film from n-doping to p-doping to underlying SWCNT contact sections. This is seen by comparing the current in Fig.\ 4d for p and n doping of the channel at $V_{G}-V_{CNP} = \pm$ 6 V to the left (holes) or right (electrons) of the charge degeneracy point at $V_{CNP} = 9$ V. The conductance is an order of magnitude larger under p-doping of the channel than n-doping. For p-doping, small remnants of conductance oscillations suggest the formation of an open QD. This is consistent with an $E_{C}^{h}$ not much larger than the thermal energy $\sim 25$ meV. On the other hand the near zero conductivity for electron doping suggests $E_{C}^{e} >>$ 25 meV. This large $e-h$ transport asymmetry, with an on-off ratio exceeding 4000, demonstrates the potential to developed two-in-one room-temperature SWCNT transistors.

\begin{figure}
\includegraphics[width=3.25in]{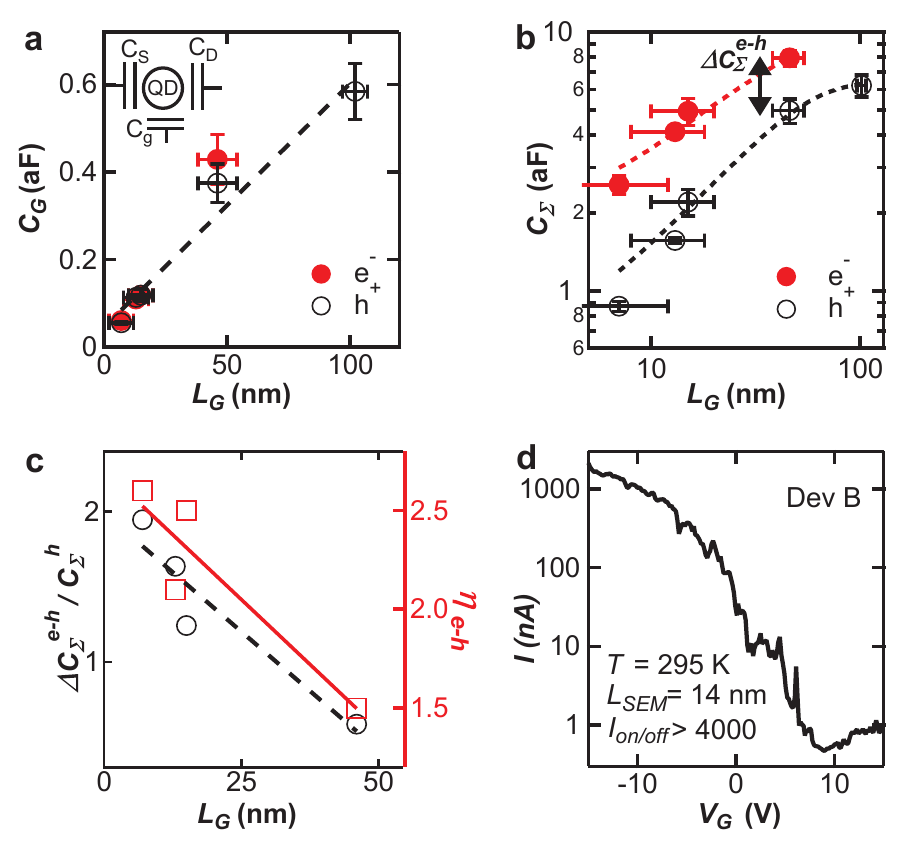}
\caption{\label{}Origin of the $e-h$ charging energy asymmetry. \textbf{a}, The gate-to-QD capacitance, $C_{G}$, in our devices is $e-h$ symmetric and scales linearly with $L_{G}$. The inset cartoon of the QD devices shows the lumped element interpretation of $C_{G}$, $C_{S}$, and $C_{D}$. \textbf{b}, The total QD capacitance $C_{\Sigma}$ increases approximately linearly as a function of $L_{G}$ in our devices. A clear capacitance offset $\Delta C^{e-h}_{\Sigma}$ is visible between the hole and electron data due to the $e-h$ tunnel barrier asymmetry. This offset is at the origin of the $e-h$ charging energy asymmetry. \textbf{c} The relative $e-h$ capacitance asymmetry $\Delta C^{e-h}_{\Sigma} / C^{h}_{\Sigma}$ decreases with $L_{G}$, and explains why $\eta_{e-h} = E_{C}^{h}/E_{C}^{e}$ also decreases with $L_{G}$. The dashed and solid lines are linear fits of the data. \textbf{d}, Room-temperature $I-V_{G}$ transport characteristics in Device B ($L_{SEM}=$ 14 nm). When comparing $I$ at $V_{G}-V_{CNP}=\pm$ 6 V, the data show a clear $e-h$ asymmetry with a conductance over an order of magnitude larger when the channel is hole-doped compared to electron-doped.}
\end{figure}

\subsection{Conclusion}

In summary, we used suspended annealed gold films as local gates to create $e-h$ asymmetric suspended SWCNT-QD transistors (Fig.\ 1e). With transport measurements, we showed that the gold gates permit ballistic charge transport in the underlying SWCNT sections, and create sharp tunnel barriers at the edge of the SWCNT channel. These barriers were of vastly different heights whether the channel is hole or electron doped. This produced a giant $e-h$ transport asymmetry in the five SWCNT transistors studied. We used this asymmetry to create two new types of two-in-one SWCNT quantum devices. We first showed in a low-$E_{g}$ SWCNT device that a small gate voltage could switch the device from being a 330 nm long quantum bus under hole doping to a 102 nm long QD for electron doping. Secondly, we reported four devices with $E_{g} \gtrsim$ 200 meV where a small $V_{G}$ could switch the QD between two charging energies whose values differed by a factor up to 2.6. We established that the amount of $e-h$ transport asymmetry in these devices scales inversely with length of the channel and the band gap of the SWCNT. In a 14-nm long channel device, we observed that this $e-h$ asymmetry survives at room temperature. Nanotube transistors with a giant $e-h$ transport asymmetry could find applications in exploring the physics of ultra-short quantum SWCNT-NEMS, to shrink down the size of SWCNT qubits, and to add functionalities to room temperature SWCNT transitors.

\subsection{Methods}
Lithography. We use e-beam lithography to define bow-tie shaped breakjunctions in a PMMA bilayer on SWCNTs deposited on a SiO$_{2}$/Si substrate. We then evaporate a 40-nm thick gold film (no adhesion layer) to define the gold junctions. We use a wet buffered oxide etch to freely suspend the central portion of the gold breakjunctions as shown in Figure 1a. \newline
Electromigration. The red data in Fig.\ 1b show the first stage of EM where we narrow down the central bowtie-shaped gold junctions. To do so, we ramp up a bias voltage, $V_{B}$, across the device and monitor the current, $I$. As the resistance, $R= V_{B}/I$, increases a feedback circuit rapidly ramps down $V_{B}$ to avoid an avalanche breaking of the junction. We repeat the feedback controlled-EM process iteratively to gradually narrow down the junction (increase $R$), and these successive voltage ramps are visible in the inset of Fig 1b. We learned previously that using a continuous $V_{B}$ ramp to break a gold junction gives a central junction gap which is proportional to the initial resistance value of the junction \cite{Island11, Island12, Tayari15}. Thus we stop the narrowing of the gold junction at a desired initial resistance in order to achieve a target channel length. We then proceed to step two of the EM (blue data), where the junction is broken with a single continuous voltage ramp. \newline
Band Gap Measurement. To determine the band gaps of the tubes in our devices, we use the data at the charge neutrality point $N =$ 0, corresponding to the largest blockade diamond in Figs. 2c, 3c-e, and S3. The band gap is given by $E_{g} = E_{add}^{N=0} - E_{C}^{N=0}- \Delta$ where $E_{add}^{N=0}$ is the height of the $N =$ 0 diamond, $E_{C}$ is the charging energy, and $\Delta$ is the single particle energy spacing. See SI section 2 for details.

\subsection{Supplementary Information}
Supplementary information is available in the online version of the paper. Reprints and
permissions information are available online at . Correspondence and requests for materials should be addressed to A.R.C.
\subsection{Acknowledgements}
This work was supported by NSERC (Canada), CFI (Canada), and Concordia University. We acknowledge usage of the QNI (Quebec Nano Infrastructure) cleanroom network.

\subsection{Author Contributions}
A.C.M. led all steps of the fabrication of the samples and acquired all the data reported. A.R.C. conceived and supervised the experiment. A.C.M and A.R.C. did the data analysis. V.T. and J.M.P. contributed to the sample fabrication and the development of the experimental methods. A.R.C. and A.C.M. wrote the manuscript. All authors discussed the results and the manuscript.

\end{document}